\documentclass[12pt]{article}
\usepackage{axodraw,bbold}

\catcode`@=12
\begin{document}
\thispagestyle{empty}
\begin{flushright} 
UCRHEP-T453\\ 
August 2008\
\end{flushright}
\vspace{0.5in}
\begin{center}
{\LARGE	\bf Dark Scalar Doublets and Neutrino\\ Tribimaximal Mixing from 
A$_4$ Symmetry\\}
\vspace{1.5in}
{\bf Ernest Ma\\}
\vspace{0.2in}
{\sl Department of Physics and Astronomy, University of 
California,\\ Riverside, California 92521, USA\\}
\vspace{1.5in}
\end{center}

\begin{abstract}\
In the context of $A_4$ symmetry, neutrino tribimaximal mixing is achieved 
through the breaking of $A_4$ to $Z_3$ ($Z_2$) in the charged-lepton 
(neutrino) sector respectively.  The implied vacuum misalignment of the  
(1,1,1) and (1,0,0) directions in $A_4$ space is a difficult technical 
problem, and cannot be treated without many auxiliary fields and symmetries 
(and perhaps extra dimensions).  It is pointed out here that an alternative 
scenario exists with $A_4$ alone and no redundant fields, if neutrino masses 
are ``scotogenic'', i.e. radiatively induced by dark scalar doublets as 
recently proposed.

\end{abstract}

\newpage
\baselineskip 24pt

The neutrino mixing angles are now known to some accuracy.  Based on a recent 
global analysis \cite{gm08},
\begin{equation}
\theta_{23} = 42.3~(+5.1/-3.3), ~~~ \theta_{12} = 34.5 \pm 1.4, ~~~ \theta_{13} 
= 0.0~(+7.9/-0.0),
\end{equation}
at the $1\sigma$ level.  Thus the central values of $\sin^2 2 \theta_{23}$, 
$\tan^2 \theta_{12}$, and $\theta_{13}$ are 0.99, 0.47, and 0 respectively.  
These numbers agree well with the hypothesis of tribimaximal mixing 
\cite{hps02}, i.e.
\begin{equation}
\sin^2 2 \theta_{23} = 1, ~~~ \tan^2 \theta_{12} = 0.5, ~~~ \theta_{13} = 0.
\end{equation}
Such a pattern is best understood as the result of a family symmetry and 
the non-Abelian finite group $A_4$ has proved to be useful in this 
regard \cite{mr01,m02,bmv03}.  Specifically, it was shown 
\cite{m04,af05,bh05} how this may be achieved by the breaking of $A_4$ 
in a prescribed manner \cite{m06}, i.e. $A_4 \to Z_3$ in the charged-lepton 
sector and $A_4 \to Z_2$ in the neutrino sector.  The group-theoretical 
framework of how this works in general has also been discussed 
\cite{l07,bhl08}.  For a brief history, see Ref.~\cite{m07}.

In another development, it has been proposed recently \cite{m06-1} that 
neutrino mass is radiative in origin such that the particles in the loop 
are odd under a new discrete $Z'_2$ symmetry, thereby accommodating 
a dark-matter candidate.  The simplest realization  
of this ``scotogenic'' neutrino mass is depicted in Fig.~1.
\begin{figure}[htb]
\begin{center}
\begin{picture}(360,120)(0,0)
\ArrowLine(90,10)(130,10)
\ArrowLine(180,10)(130,10)
\ArrowLine(180,10)(230,10)
\ArrowLine(270,10)(230,10)
\DashArrowLine(155,85)(180,60)3
\DashArrowLine(205,85)(180,60)3
\DashArrowArc(180,10)(50,90,180)3
\DashArrowArcn(180,10)(50,90,0)3
\Text(110,0)[]{$\nu_i$}
\Text(250,0)[]{$\nu_j$}
\Text(180,0)[]{$N_k$}
\Text(135,50)[]{$\eta^0$}
\Text(230,50)[]{$\eta^0$}
\Text(150,90)[]{$\phi^{0}$}
\Text(217,90)[]{$\phi^{0}$}
\end{picture}
\end{center}
\caption{One-loop generation of seesaw neutrino mass.}
\end{figure}
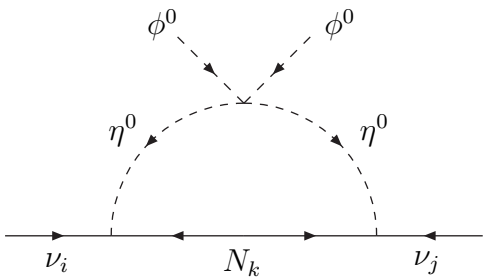
Here $N_k$ are heavy Majorana fermion singlets odd under $Z'_2$ and 
$(\eta^+,\eta^0)$ is a scalar doublet also odd under $Z'_2$ \cite{dm78}, 
whereas the standard-model $(\phi^+,\phi^0)$ is even.  Exact conservation of 
$Z'_2$ means of course that $\eta^0$ has no vacuum expectation value, so that 
$N$ is not the Dirac mass partner of $\nu$ as usually assumed.  The 
allowed quartic coupling $(\lambda_5/2)(\Phi^\dagger \eta)^2 + H.c.$ 
splits Re($\eta^0$) and Im($\eta^0$) so that whichever is lighter is a good 
dark-matter candidate \cite{m06-1,bhr06,lnot07,glbe07}.  The collider 
signatures of Re($\eta^0$) and Im($\eta^0$) have also been discussed 
\cite{cmr07}.  For a brief review of the further developments of this idea, 
see Ref.~\cite{m08}.

Going back to $A_4$, let $(\nu_i,l_i) \sim \underline{3}$ and either (I) 
$l_i^c \sim \underline{1}, \underline{1}', \underline{1}''$, or (II) 
$l_i^c \sim \underline{3}$, then with the Higgs fields (I) $(\phi^+_i,
\phi^0_i) \sim \underline{3}$, or (II) $(\phi^+_i,\phi^0_i) \sim 
\underline{3}$ and $(\zeta^+,\zeta^0) \sim \underline{1}$, the mass 
matrix linking $l_i$ to $l^c_j$ is diagonalized on the left by \cite{m06}
\begin{equation}
U_{l\nu} = {1 \over \sqrt{3}} \pmatrix{1 & 1 & 1 \cr 1 & \omega & \omega^2 \cr 
1 & \omega^2 & \omega},
\end{equation}
where $\omega = \exp(2\pi i/3) = -1/2 + i \sqrt{3}/2$, if $\langle \phi^0_1 
\rangle = \langle \phi^0_2 \rangle = \langle \phi^0_3 \rangle = v$.  This 
is a natural minimum of the Higgs potential \cite{mr01} because it 
corresponds to a $Z_3$ residual symmetry with $e \sim 1$, $\mu \sim \omega^2$, 
$\tau \sim \omega$, whereas $\Phi \equiv (\Phi_1 + \Phi_2 + \Phi_3)/\sqrt{3} 
\sim 1$, $\Phi' \equiv (\Phi_1 + \omega \Phi_2 + \omega^2 \Phi_3)/\sqrt{3} 
\sim \omega^2$, and $\Phi'' \equiv (\Phi_1 + \omega^2 \Phi_2 + \omega \Phi_3)/
\sqrt{3} \sim \omega$.

To obtain tribimaximal mixing, what is required for the Majorana neutrino 
mass matrix ${\cal M}_\nu$ is \cite{m04} $2-3$ symmetry and zero $1-2$ and 
$1-3$ entries.  Since $123 + 231 + 312$ and $132 + 321 + 213$ are $A_4$ 
invariants and ${\cal M}_\nu$ must be symmetric, the simplest implementation 
is to have \cite{af05}
\begin{equation}
{\cal M}_\nu = \pmatrix{a & 0 & 0 \cr 0 & a & d \cr 0 & d & a},
\end{equation}
which requires effective scalar triplet fields $(\xi^{++}_i,\xi^+_i,\xi^0_i)$ 
transforming as \underline{3} with $\langle \xi^0_1 \rangle \neq 0$ and 
$\langle \xi^0_{2,3} \rangle = 0$, thereby breaking $A_4 \to Z_2$.  Let the 
eigenvalues of ${\cal M}_\nu$ be denoted by
\begin{equation}
m_1=a+d, ~~~ m_2=a, ~~~ m_3=-a+d,
\end{equation}
then the mixing matrix linking $\nu_{e,\mu,\tau}$ to $\nu_{1,2,3}$ is given 
by \cite{m07}
\begin{equation}
(U_{l\nu})^\dagger \pmatrix{1 & 0 & 0 \cr 0 & 1/\sqrt{2} & -1/\sqrt{2} \cr 
0 & 1/\sqrt{2} & 1/\sqrt{2}} \pmatrix{0 & 1 & 0 \cr 1 & 0 & 0 \cr 0 & 0 & i} 
= \pmatrix{\sqrt{2/3} & 1/\sqrt{3} & 0 \cr -1/\sqrt{6} & 1/\sqrt{3} & 
-1/\sqrt{2} \cr -1/\sqrt{6} & 1/\sqrt{3} & 1/\sqrt{2}},
\end{equation}
i.e. tribimaximal mixing.

Because the scalar fields $\Phi_i$ and $\xi_i$ are both \underline{3} under 
$A_4$, the requirement that they break the vacuum in different directions 
is incompatible with the most general Higgs potential allowed by $A_4$ 
alone.  Complicated sets of auxiliary fields and symmetries (and/or possible 
extra dimensions) are then needed \cite{af05,bh05,af06,h07,cdgg08} for it to 
happen.  This is perhaps the one stumbling block of the application of $A_4$ 
to tribimaximal mixing.

The reason that the two breaking directions are incompatible is because 
$A_4$ allows $\underline{3} \times \underline{3}$ to be invariant, 
so if one \underline{3} has a vacuum expectation value along a certain 
direction, the other is forced to as well.  This is of course not a 
problem if only one \underline{3} is required to have vacuum expectation 
values and not the other, because that corresponds to having an exactly 
conserved $Z'_2$ under which the second \underline{3} is odd.  Specifically, 
let the charged leptons acquire mass from $\Phi_i$, but the neutrino masses 
are obtained radiatively as discussed earlier, without any vacuum expectation 
value for $\eta^0$.

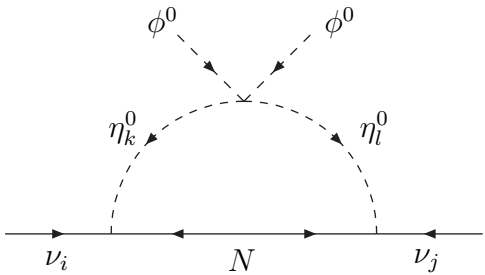
\begin{figure}[htb]
\begin{center}
\begin{picture}(360,120)(0,0)
\ArrowLine(90,10)(130,10)
\ArrowLine(180,10)(130,10)
\ArrowLine(180,10)(230,10)
\ArrowLine(270,10)(230,10)
\DashArrowLine(155,85)(180,60)3
\DashArrowLine(205,85)(180,60)3
\DashArrowArc(180,10)(50,90,180)3
\DashArrowArcn(180,10)(50,90,0)3
\Text(110,0)[]{$\nu_i$}
\Text(250,0)[]{$\nu_j$}
\Text(180,0)[]{$N$}
\Text(135,50)[]{$\eta^0_k$}
\Text(230,50)[]{$\eta^0_l$}
\Text(150,90)[]{$\phi^{0}$}
\Text(217,90)[]{$\phi^{0}$}
\end{picture}
\end{center}
\caption{One-loop generation of seesaw neutrino mass.}
\end{figure}

Instead of having three $N$'s (which would have been necessary in the 
canonical seesaw mechanism), assume just one $N$ but three scalar $\eta$ 
doublets, as shown in Fig.~2. 
Let $(\eta^+_i,\eta^0_i)$ transform as \underline{3} under $A_4$, 
then ${\cal M}_\nu$ is proportional to the unit matrix.  Suppose $A_4$ is now 
softly broken by the quadratic terms $\eta_2^\dagger \eta_3 + \eta_3^\dagger 
\eta_2$ and $2 \eta_1^\dagger \eta_1 - \eta_2^\dagger \eta_2 - \eta_3^\dagger 
\eta_3$.  Then ${\cal M}_\nu$ is of the form
\begin{equation}
{\cal M}_\nu = \pmatrix{a+2b & 0 & 0 \cr 0 & a-b & d \cr 0 & d & a-b},
\end{equation}
which will lead to tribimaximal mixing \cite{m04}, with
\begin{equation}
m_1=a-b+d, ~~~ m_2=a+2b, ~~~ m_3=-a+b+d.
\end{equation}
Since the origin of ${\cal M}_\nu$ is the mass-squared matrix of 
$\eta^0_{1,2,3}$, this model may be tested at least in principle.  Note 
that $b=0$ cannot be a solution here as in Ref.~\cite{af05} because that 
would require a negative mass-squared eigenvalue for $\eta^0_i$.  As it is, 
$\Delta m^2_{sol} << \Delta m^2_{atm}$ implies $d \simeq 3b$ or $-2a-b$ 
in this scenario.

Consider now the scalar sector in more detail.  Since $\eta_i$ are odd under 
the new exactly conserved $Z'_2$ for dark matter, and have no vacuum 
expectation value.  The bilinear terms $\Phi_i^\dagger \eta_j$ are forbidden, 
and the quartic terms must contain an even number of $\Phi_i$ and $\eta_j$. 
The scalar potential consisting of only $\Phi_i$ is given by \cite{mr01}
\begin{eqnarray}
V_\Phi &=& m^2 \sum_i \Phi_i^\dagger \Phi_i + {1 \over 2} \lambda_1 \left( 
\sum_i \Phi_i^\dagger \Phi_i \right)^2 \nonumber \\
&+& \lambda_2 (\Phi_1^\dagger \Phi_1 + \omega^2 \Phi_2^\dagger \Phi_2 + 
\omega \Phi_3^\dagger \Phi_3)(\Phi_1^\dagger \Phi_1 + \omega \Phi_2^\dagger 
\Phi_2 + \omega^2 \Phi_3^\dagger \Phi_3) \nonumber \\ 
&+& \lambda_3 [(\Phi_2^\dagger \Phi_3)(\Phi_3^\dagger \Phi_2) + (\Phi_3^\dagger 
\Phi_1)(\Phi_1^\dagger \Phi_3) + (\Phi_1^\dagger \Phi_2)(\Phi_2^\dagger \Phi_1) 
\nonumber \\
&+& \left\{ {1 \over 2} \lambda_4 [(\Phi_2^\dagger \Phi_3)^2 + (\Phi_3^\dagger 
\Phi_1)^2 + (\Phi_1^\dagger \Phi_2)^2] + H.c. \right\}.  
\end{eqnarray}
The parameters $m^2$ and $\lambda_{1,2,3}$ are automatically real, and 
$\lambda_4$ may be chosen real by rotating the overall phase of $\Phi_i$. 
The vacuum solution
\begin{equation}
v_1=v_2=v_3=v=[-m^2/(3\lambda_1+2\lambda_3+2\lambda_4)]^{1/2}
\end{equation}
is protected by the residual symmetry $Z_3$, under which
\begin{eqnarray}
\Phi &\equiv& {1 \over \sqrt{3}} (\Phi_1 + \Phi_2 + \Phi_3) ~\sim~ 1 \\
\Phi' &\equiv& {1 \over \sqrt{3}} (\Phi_1 + \omega \Phi_2 + \omega^2 \Phi_3) 
~\sim~ \omega^2, \\
\Phi'' &\equiv& {1 \over \sqrt{3}} (\Phi_1 + \omega^2 \Phi_2 + \omega \Phi_3) 
~\sim~ \omega, 
\end{eqnarray}
as already mentioned. The scalar doublet $\Phi$ has the properties of the 
standard-model Higgs doublet with mass-squared eigenvalues 
$2(3\lambda_1+2\lambda_3+\lambda_4)v^2$, 0, and 0 for $\sqrt{2}$Re$\phi^0$, 
$\sqrt{2}$Im$\phi^0$, and $\phi^{\pm}$ respectively.  The charged scalars 
${\phi'}^\pm$ and ${\phi''}^\pm$ have $m^2_\pm = -3(\lambda_3+\lambda_4)v^2$, 
whereas ${\phi'}^0$ and ${\phi''}^0$ are not mass eigenstates, but rather 
${\phi'}^0 = (\psi_1 + \psi_2)/\sqrt{2}$ and ${\phi''}^0 = (\psi_1^* - 
\psi_2^*)/\sqrt{2}$, i.e.
\begin{eqnarray}
\psi_1 &=& {1 \over \sqrt{2}} {\rm Re}({\phi'}^0 + {\phi''}^0) + 
{i \over \sqrt{2}} {\rm Im}({\phi'}^0 - {\phi''}^0) ~\sim~ \omega^2, \\
\psi_2 &=& {1 \over \sqrt{2}} {\rm Re}({\phi'}^0 - {\phi''}^0) + 
{i \over \sqrt{2}} {\rm Im}({\phi'}^0 + {\phi''}^0) ~\sim~ \omega^2,
\end{eqnarray}
with $m_1^2 = 2(3\lambda_2-\lambda_3-\lambda_4)v^2$ and $m_2^2 = -6\lambda_4 
v^2$.  This subtlety in the mass spectrum of ${\phi'}^0$ and ${\phi''}^0$ 
was not recognized in Ref.~\cite{mr01}, where $\tau^- \to \mu^- \mu^+ e^-$ 
and $\mu \to e \gamma$ were thought to be nonzero.  In fact, they are 
forbidden by the residual $Z_3$ symmetry.

The addition of $\eta_i$ to the scalar potential does not change the above 
because $\langle \eta^0_i \rangle = 0$ and $Z'_2$ remains exactly conserved. 
However, the breaking of $A_4 \to Z_3$ by $\langle \phi_i^0 \rangle$ 
generates additional contributions to the $\eta_i^0$ mass-squared matrix 
of the form
\begin{eqnarray}
&& \Delta_1^2 (\eta_1^* \eta_1 + \eta_2^* \eta_2 + \eta_3^* \eta_3) + 
\{ \Delta_2^2 (\eta_1^* \eta_2 + \eta_2^* \eta_3 + \eta_3^* \eta_1) + c.c. \} 
\nonumber \\
&& + \{ {1 \over 2} \Delta_3^2 (\eta_1^2 + \eta_2^2 + \eta_3^2) + c.c. \} + 
\{ \Delta_4^2 (\eta_1 \eta_2 + \eta_2 \eta_3 + \eta_3 \eta_1) + c.c. \}.
\end{eqnarray}
In other words, except for soft terms, the complete Higgs potential remains 
invariant under $Z_3$ after spontaneous symmetry breaking.  The induced 
neutrino mass matrix of Eq.~(7) is then modified:
\begin{equation}
{\cal M}_\nu = \pmatrix{a+2b & e & e \cr e & a-b & d \cr e & d & a-b}.
\end{equation}
Since the one-loop neutrino mass of Fig.~1 is proportional to $\Delta_3^2$ 
and $\Delta_4^2$ which split Re($\eta_i^0$) and Im($\eta_i^0$), these 
parameters should be relatively small.  Assuming that $\Delta_2^2$ is 
also small, then $e$ should be small compared to $a,b,d$ in Eq.~(17). 
This means that \cite{m02-1} $\sin^2 2 \theta_{23} = 1$ and $\theta_{13} = 0$ 
as before, but the solar mixing angle is now given by
\begin{equation}
\tan^2 \theta_{12} = {1 \over 2} (1- 6 \epsilon + 15 \epsilon^2),
\end{equation}
where $\epsilon = e/(d-3b)$.  Thus $\tan^2 \theta_{12} = 0.47$ is obtained 
for $\epsilon = 0.01$.

One possible explanation of the smallness of the terms in Eq.~(16) is 
that $\Phi$ and $\eta$ are separated in an extra dimension so that they 
communicate only through a singlet in the bulk.  In the limit this effect 
vanishes, there would be no mass splitting between Re($\eta^0$) and 
Im($\eta^0$), resulting in zero neutrino mass and no viable dark-matter 
candidate. With it, neutrinos acquire small radiative Majorana seesaw 
masses, Re($\eta^0$) is a good dark-matter candidate, and near tribimaximal 
mixing is possible.

In conclusion, it has been shown how $A_4$ symmetry may be implemented 
in a model of ``scotogenic'' neutrino mass with dark scalar doublets. 
The neutrino mass matrix is induced by the neutral scalar mass-squared 
matrix spanning Re($\eta^0_{1,2,3}$) and Im($\eta^0_{1,2,3}$).  This scheme 
allows the neutrino mixing angles $\theta_{23}$ and $\theta_{13}$ to be 
exactly $\pi/4$ and 0, whereas $\tan^2 \theta_{12}$ should not be exactly 
1/2.  Suppose the lightest Re($\eta^0$) is dark matter, then its possible 
discovery \cite{cmr07} at the LHC together with the other $\eta$ particles 
in accordance with Fig.~2 would be a verifiable test of this proposal.

This work was supported in part by the U.~S.~Department of Energy under Grant 
No.~DE-FG03-94ER40837.

%\newpage

\baselineskip 16pt

\bibliographystyle{unsrt}

\end{document}